\documentclass[journal=nalefd,manuscript=letter]{achemso}
\usepackage[version=3]{mhchem} 
\usepackage{mathrsfs}
\usepackage{graphicx}
\usepackage{dcolumn}
\usepackage{bm}
\usepackage{color}
\usepackage{float}

\title{Tunable Dot Platform for Controlling Electron Flow in Graphene}

\author{Fereshte Ildarabadi}
\affiliation{School of Physical Sciences, Dublin City University, Glasnevin, Dublin 9, Ireland}
\author{Stephen R. Power}
\email{stephen.r.power@dcu.ie}
\affiliation{School of Physical Sciences, Dublin City University, Glasnevin, Dublin 9, Ireland}

\date{\today}

\begin{document}


\begin{abstract}
We introduce an innovative graphene-based architecture to control electronic current flows. 
The tunable dot platform (TDP) consists of an array of gated dots, with independently adjustable potentials, embedded in graphene. 
Inspired by Mie theory, and leveraging multiscattering effects, we demonstrate that tailored current behavior can be achieved due to the variety of possible dot configurations.  
Optimization is performed using differential evolution, which identifies configurations that maximize specific objectives, such as directing or splitting an electron beam by tuning the angular dependence of scattering. 
Our results demonstrate the potential of the TDP to provide precise control over induced current flows in graphene, making it a promising component for next-generation electronic and electron optic devices. 
\end{abstract}

\maketitle

\paragraph{Introduction.} Quantum transport in graphene, governed by massless Dirac fermions, offers exceptional potential for ultrafast electronics beyond conventional semiconductors \cite{castro2009electronic, you2020laser, xu2024graphene, wang2021graphene, zhu2017graphene, geim2007rise}. 
Its gapless conical energy dispersion gives rise to chirality, which suppresses backscattering \cite{ando1998berry, suzuura2002crossover} and enables Klein tunneling so that electrons can traverse potential barriers regardless of their height or width, as confirmed by Fabry-Pérot interference and resistance measurements \cite{katsnelson2006chiral, allain2011klein, young2009quantum, stander2009evidence}. 
Although this makes graphene ideal for high-speed transport, it hinders the charge confinement essential for many device applications. 
Various strategies have been proposed to overcome this limitation, for example, periodic potentials \cite{park2008anisotropic, park2008new, forsythe2018band}, perforations \cite{pedersen2008GALs, power2012GALs, jessen2019lithographic} or the application of electric fields in bilayer systems \cite{zhang2009direct} to induce band gaps.

An alternative approach is to directly leverage the ballistic, wave-like nature of Dirac fermions in graphene. 
Instead of relying on bandgap engineering, electron optic devices~\cite{chakraborti2024electron,ren2024electron,yu2022electron,lagasse2020unveiling,cheianov2007focusing, young2009quantum, chen2016electron,calogero2018electron, wang2019graphene} use electrostatic gates to steer carriers, similar to light manipulation in optics. 
Unlike conventional transistors that control current via carrier accumulation or depletion, this approach enables precise trajectory control through collimation at potential interfaces. 
Local gating~\cite{ingla2023specular,cheianov2007focusing, sajjad2011high} allows for such control without large-scale structures. 
Recent advances in electron metasurfaces show that phase-engineered gating can shape electron wavefronts at room temperature~\cite{zhao2023electron}. 
Functional electron optic devices such as Dirac fermion microscopes, waveguides, and beam splitters have also been proposed or demonstrated~\cite{boggild2017two, forrester2023electron, brandimarte2017tunable, barnard2017absorptive, rickhaus2015guiding}.
  
Quantum dots (QDs), in the form of circular potential steps, offer a rich framework not only for confining electrons in graphene systems, but also for controlling the flow of current in their vicinity~\cite{gutierrez2016klein,bai2018generating,heinisch2013mie, abdullah2018graphene,lee2016imaging, matulis2008quasibound,jing2022gate}.
A QD acts as an electronic lens for propagating  electrons, drawing an analogy to Mie scattering in optics, where light interacts with small particles \cite{heinisch2013mie}. 
QDs with broken inversion symmetry further introduce valley-dependent scattering properties~\cite{aktor2021valley, solomon2021valley}.
In all cases, the scattering behavior is strongly influenced by the size and strength of the QD, as well as the incident electron energy. 
Beyond individual dots, multiple electron scattering in graphene has been considered for various QD configurations~\cite{vaishnav2011intravalley, walls2016talbot, ren2019effective, sadrara2019dirac, sadrara2022collective}, including metasurfaces and bias-tunable metagratings. 
These assemblies allow electron flow to be finely controlled and enable tailored electron trajectories and interference patterns.

\begin{figure}
	\centering
	\includegraphics[scale=0.55]{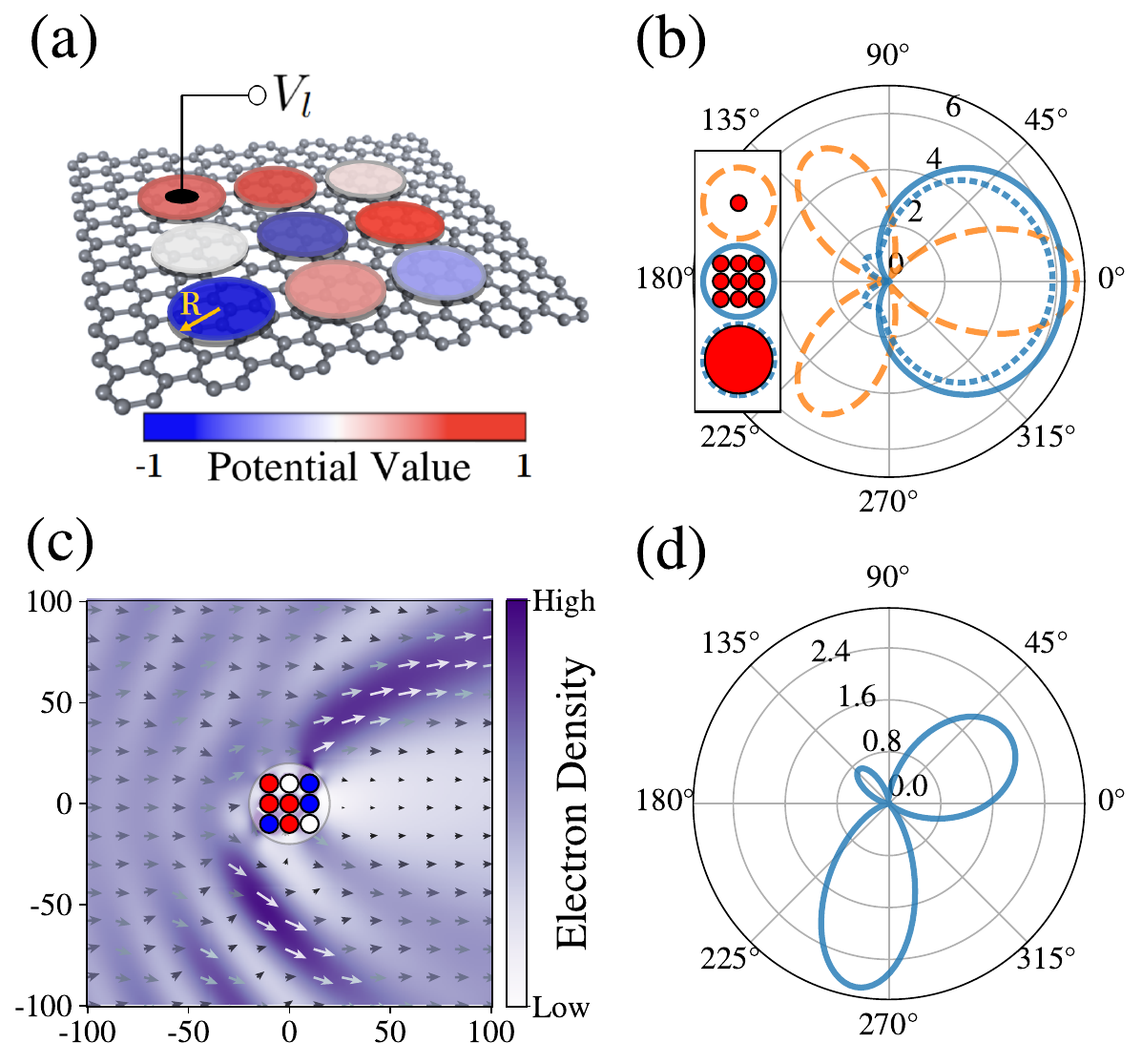}
	\caption{ (a) Schematic of the TDP architecture embedded in graphene. (b) Far-field angular dependence of scattering current ($rj^{\text{sc},r}/R$) shown in polar coordinates for a single QD (dashed), a simple TDP with $R=4.5$ (solid), and a large effective QD with $R_{eff}=19.2$ (dotted), all with potential $V = 1$ (red dots). (c) Electron density (background color) and current flow (arrows) for a TDP with a random configuration of dot potentials $V = 0, \pm 1$. (d) Far-field scattering current for the TDP in (c). All results are at energy $E = 0.1$.} 
    \label{fig1}
\end{figure}

Motivated by the use of QDs to fine-tune electron dynamics, we introduce a new architecture, the Tunable Dot Platform (TDP), to engineer angle-dependent electron flows in graphene. 
The TDP, shown in Fig.~\ref{fig1}(a), consists of a $3\times3$ matrix of QDs embedded in graphene.
While the position and size of the dots are fixed, the potential of each dot can be independently tuned, enabling a wide variety of dot configurations. 

Using generalized Mie theory \cite{mie1908beitrage} to simulate multiple scattering within the TDP, we show that this platform allows a diverse range of position and angle-dependent current profiles to be generated.
Specific behaviors observed for individual dots can be reinforced or suppressed as required, while the symmetry breaking supported by TDPs allows for new transport features that are not possible with single dots.

A key feature of the TDP is programmability: the potentials on individual dots can be tuned by different gates, similar to setups already used to investigate qubits in graphene QDs \cite{banszerus2018gate,freitag2016electrostatically,recher2010quantum,ge2020visualization}. 
This will enable the platform to be reconfigured in-situ to the optimal arrangement required for a specific outcome. 
To find such optimal TDP configurations, we employ a differential evolution (DE) approach~\cite{storn1997differential,price2006differential,chakraborty2008advances}. 
This strategy, inspired by biological evolution, iteratively refines dot configurations to maximize a defined fitness function. In our case, fitness is determined by the ability of the TDP to achieve particular electron optic behaviors, such as directing electron flow at specific angles.

We first outline the theoretical framework for single and multiple dot scattering, before briefly summarizing the scattering properties of single QDs. 
The scattering characteristics of TDP systems are then discussed, before we explain the DE algorithm for optimizing TDP configurations and demonstrate results for different far-field targets. 
The approach is then generalized to allow the optimization of near-field currents, before we end by discussing possible experimental implementations and extensions of the platform.

\paragraph{Scattering Theory.} \label{method}

We begin with a single QD in graphene and address the electron scattering problem using the single-valley Dirac Hamiltonian at low energies:
\begin{eqnarray}
\mathscr{H}(k) = \hbar v_F \begin{pmatrix}
V\Theta(R-r) & k e^{-i\phi} \\
k e^{i\phi} & V\Theta(R-r)
\end{pmatrix},\nonumber
\end{eqnarray}
where the Heaviside function $\Theta(R-r)$ confines the potential $V \neq 0$ to inside a circular dot of radius $R$. 
Due to the rotational symmetry of the dots the wavenumber $\bm{k} = (k_x, k_y)$ is expressed in polar coordinates. 
In the following, we set $\hbar v_F=1$ and expand the wavefunction in each region in terms of angular momentum basis states $m = 0, \pm1, \pm2, \dots$ with a general form: 
\begin{equation}
\psi( kr, \theta) = \frac{1}{\sqrt{2}} \sum_{m=-\infty}^{\infty} C_m i^m  
\begin{pmatrix} 
h_m( kr) e^{im\theta}\\ 
i\eta  h_{m+1}(kr) e^{i(m+1)\theta}
\end{pmatrix}, 
\label{eq:psi}
\end{equation}
where $C_m$ is a mode-dependent coefficient and $h_m({k}r)$ represents either a Bessel or Hankel function.
The wavefunction outside the dot consists of an incident and a scattered component, where $k = |E|$ and $\eta = \text{sgn}(E)$. 
The incident plane wave ($\psi_{\text{inc}}$) along the $x$ direction can be expressed in the form shown in Eq.~\eqref{eq:psi} using the identity $e^{ikr \cos \theta} = \sum_{m=-\infty}^{\infty} i^m J_m(kr) e^{im\theta}$, and the scattered component $\psi_{\text{sc}}$ is described by the Hankel radial function of the first kind, $H_m^{(1)}(kr)$.
Inside the dot, the wavefunction contains a transmitted component $\psi_{\text{tr}}$, where $\eta = \text{sgn}(E - V)$, and the radial part is expressed using the Bessel function $J_m(qr)$, with $q = |E - V|$ \cite{heinisch2013mie,aktor2021valley,heinisch2015electron}. 

Imposing wavefunction continuity at the dot edge gives
\begin{equation}
\begin{aligned}
\psi_\text{inc}(kR, \theta) + \psi_{\text{sc}}(kR, \theta) 
= \psi_{\text{tr}}(qR, \theta), 
\label{eq:continuty}
\end{aligned}
\end{equation}
which can be solved for each mode $m$ separately to give the scattering ($C^{\text{sc}}_m$) and transmission ($C^{\text{tr}}_m$) coefficients. 
The electron density and probability current are given by $n = \psi^{\dagger} \psi$ and  $\bm{j} = \psi^{\dagger} \bm{\sigma} \psi$ respectively, where the Pauli vector $\bm{\sigma} = (\sigma_x, \sigma_y)$.

To quantify the angular dependence of scattering, we consider the radial component of the far-field scattered current, given by 
\begin{equation}
j^{\text{sc},r}(r \rightarrow \infty, \theta) = \frac{2\eta}{\pi kr} \sum_{m,n=-\infty}^{\infty} C^\text{sc}_m C^{\text{sc}*}_n e^{i(m-n)\theta}.
\label{eq:farf}
\end{equation}
The orange dashed curve in Fig.~\ref{fig1}(b) shows the angular scattering for a QD with $R = 4.5, V=1.0$ for an incident electron wave with energy $E = 0.1$ approaching from the left~\cite{heinisch2013mie, aktor2021valley}. 
Three preferred scattering directions are observed at $\theta = 0$ and $\theta = \pm 2\pi/3$. 
The scattering is symmetric with respect to the $x$-axis as a result of the inherent symmetry of the system, with $j^{\text{sc},r}(\theta = \pm \pi) = 0$ due to the suppression of backscattering.

Small changes in $R$, $E$ or $V$ can cause significant changes in the scattering behavior, particularly near resonant energies due to  particle trapping and interference effects~\cite{sadrara2019dirac, heinisch2013mie}.
More detail on single dot scattering is given in the Supporting Information.

\paragraph{Tunable Dot Platform.} The approach above can be extended to multiple dots by enforcing wavefunction continuity for each dot.
However, when rewriting Eq.~\eqref{eq:continuty} for the $l$-th dot of the TDP, the scattered wavefunction also includes contributions from the other dots,
\begin{equation}
\psi_{\text{sc}}(kR_l, \theta_l) = \psi_{\text{sc},l}(kR_l, \theta_l) + \sum_{j \neq l}^{N} \psi_{\text{sc},j}(kR_l, \theta_l).
\label{eq:sc}
\end{equation}
Combining the appropriate wavefunction expansions \cite{sadrara2019dirac,meng2016bounds} into the continuity equations for each dot gives a system of equations for $M$-modes and $N$ dots, forming a square coefficient matrix of size $MN \times MN$ that is solved numerically to obtain the coefficients $C^{\text{sc}}_{ml}$ and $C^{\text{tr}}_{ml}$ for $l$-th dot, from which  the full wavefunctions, electron densities and currents can be calculated. More details on this procedure are given in the Supporting Information.

The far-field scattering of a simple TDP, consisting of nine identical replicas of the single dot considered earlier, with a center-to-center separation of $d = 10$, is shown by the solid blue curve in Fig.~\ref{fig1}(b).     
It exhibits significantly different behavior to the single dot, with the TDP scattering electrons over a broader range of forward angles and suppressing the large-angle scattering noted for the single dot.
However, this result is mostly due to the increased size of the scattering region and not the internal structure of the TDP. 
This can be seen from the excellent agreement between the far-field scattering of the TDP and that of a single large dot ($R_{eff}=19.2$), shown by the dotted blue curve.

While the uniform TDP above behaves like an effective single large dot, individually tuned dots allows new scattering patterns that are not achievable with a single dot.
Figs.~\ref{fig1}(c),(d) show the scattering characteristics of a TDP with potentials chosen randomly from $V = 0, \pm 1$. 
This system gives three preferential scattering directions, but these are no longer symmetric around the $x$-axis.

Adjusting the potentials within a TDP structure allows us to arbitrarily modulate the effective shape and potential profile of the scatterer.
We will now investigate the extent to which these new degrees of freedom allow us to  manipulate electron trajectories and realize electron optic functionalities.

\paragraph{Optimization.}
For general TDP systems, we allow a broader range of potentials with steps of $\delta V = 0.1$ in $-1 \le V \le 1$.
This results in an enormous configuration space of $21^9$ possible arrangements of the $3\times3$ array, highlighting the structural flexibility of the TDP. 
However, achieving a specific target response requires the determination of the corresponding potential profile, which can be written as an optimization problem. 
To solve this, we employ the Differential Evolution (DE) algorithm -- an evolutionary method that iteratively improves candidate solutions with respect to a defined fitness function without relying on gradient information. 
DE operates through mechanisms inspired by natural evolution, including mutation, crossover and selection, enabling efficient exploration of large, complex search spaces and helping to avoid premature convergence \cite{price2006differential,storn1997differential}.  
It has demonstrated effectiveness in navigating high-dimensional optimization landscapes \cite{heller2022differential,pishchalnikov2018application,lovett2013differential,schneider2019benchmarking,saber2017performance,bor2016differential}, and has also been adopted in physics to tackle of complex computational challenges.
This includes application in diverse areas such as ultrafast optics and laser pulse characterization~\cite{gerth2019regularized} and the control of various quantum systems~\cite{zahedinejad2014evolutionary, ma2015differential, yang2019improved, mahesh2023quantum}.
Simpler genetic-algorithm techniques have been applied to optimize a range of nanostructures, including plasmonic nanoarrays \cite{forestiere2012genetically} and graphene-based THz absorbers \cite{najafi2020reliable}.

To find optimal solutions for the angular dependence of scattering within the 9-dimensional search space of the TDP structure, we define a fitness function
\begin{equation}
\mathrm{Fitness} = W \int_{\theta_{\text{in}}} j^{\text{sc},r}(\theta)  d\theta - (1-W) \int_{\theta_{\text{out}}} j^{\text{sc},r}(\theta) d\theta,
\label{eq.optim}
\end{equation}
where the first term rewards far-field scattering current $j^{\text{sc},r}(\theta)$ within a target angular range $\theta_\text{in}$ and the second term penalizes current for angles outside this range ($\theta_\text{out}$).
The weight $W$ can be used to control the relative importance of generating scattering inside and suppressing it outside the target range.
The DE algorithm iteratively improves the population’s fitness over multiple generations until a convergence criterion or a maximum number of generations is reached. For more details, see the Supporting Information.

\begin{figure}
	\centering
        \includegraphics[scale=0.65]{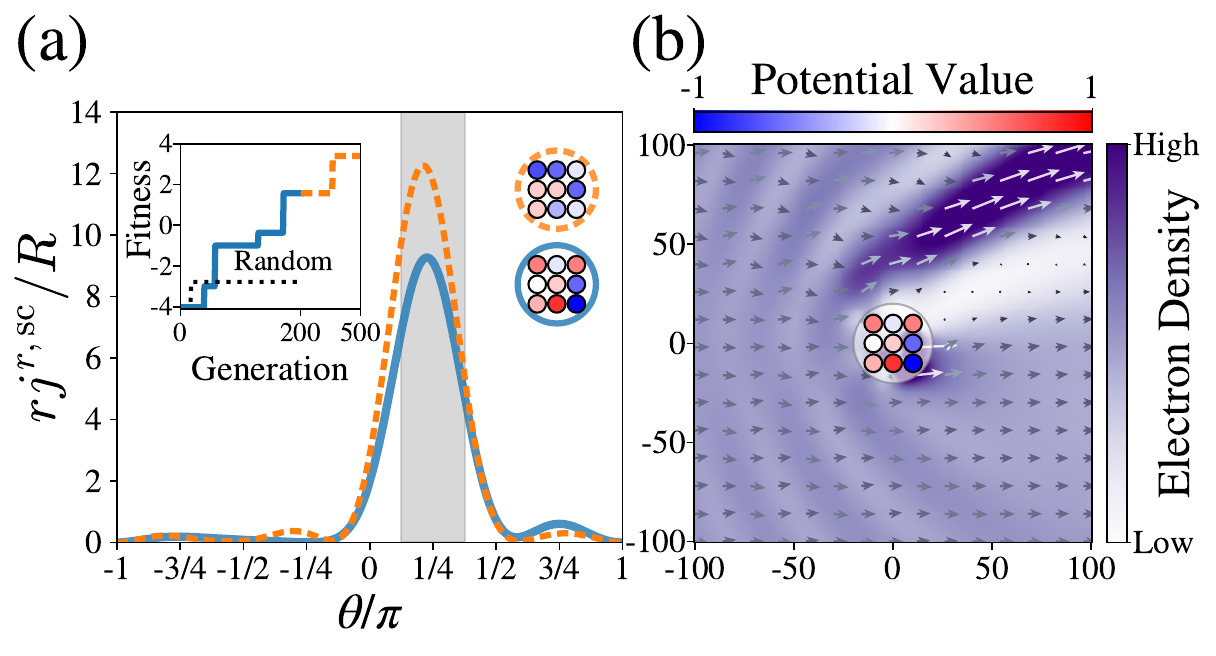}
	\caption{Optimization of far-field scattering current using DE.
(a) Far-field scattering current after 200 generations (solid curve) and 500 generations (dashed curve), with the shaded region indicating the target angles. Left inset: Fitness improvement over 200 generations (solid curve) and 500 generations (dashed curve). Right insets: Optimal TDP configurations after 200 generations (bottom) and 500 generations (top), with dot color indicating potential values.
(b) Electron density (background color) and current flow (arrows) for the optimized TDP after 200 generations, corresponding to the solid curve in (a).}  
    \label{fig2}
\end{figure}

As an example, we consider a TDP optimized to redirect far-field current into the target range $\frac{\pi}{4} \pm \frac{\pi}{8}$ shown by the shaded region in Fig.~\ref{fig2}(a). 
The resultant current after 200 generations, shown by the solid curve, has a significant peak within the target range, as desired. 
The left inset of Fig.~\ref{fig2}(a) tracks the performance of the algorithm over each generation, with the solid curve showing the best fitness value achieved so far using the DE algorithm with a population of 30 configurations. 
For comparison, the black dotted line shows the best fitness achieved by a completely random generation of new configurations in each generation. 
This highlights that the DE approach is not converging to suitable solutions by chance, but is efficiently exploring the large solution space to find the rare configurations that give the required behavior.

\begin{figure}[ht]
	\centering
	\includegraphics[scale=0.55]{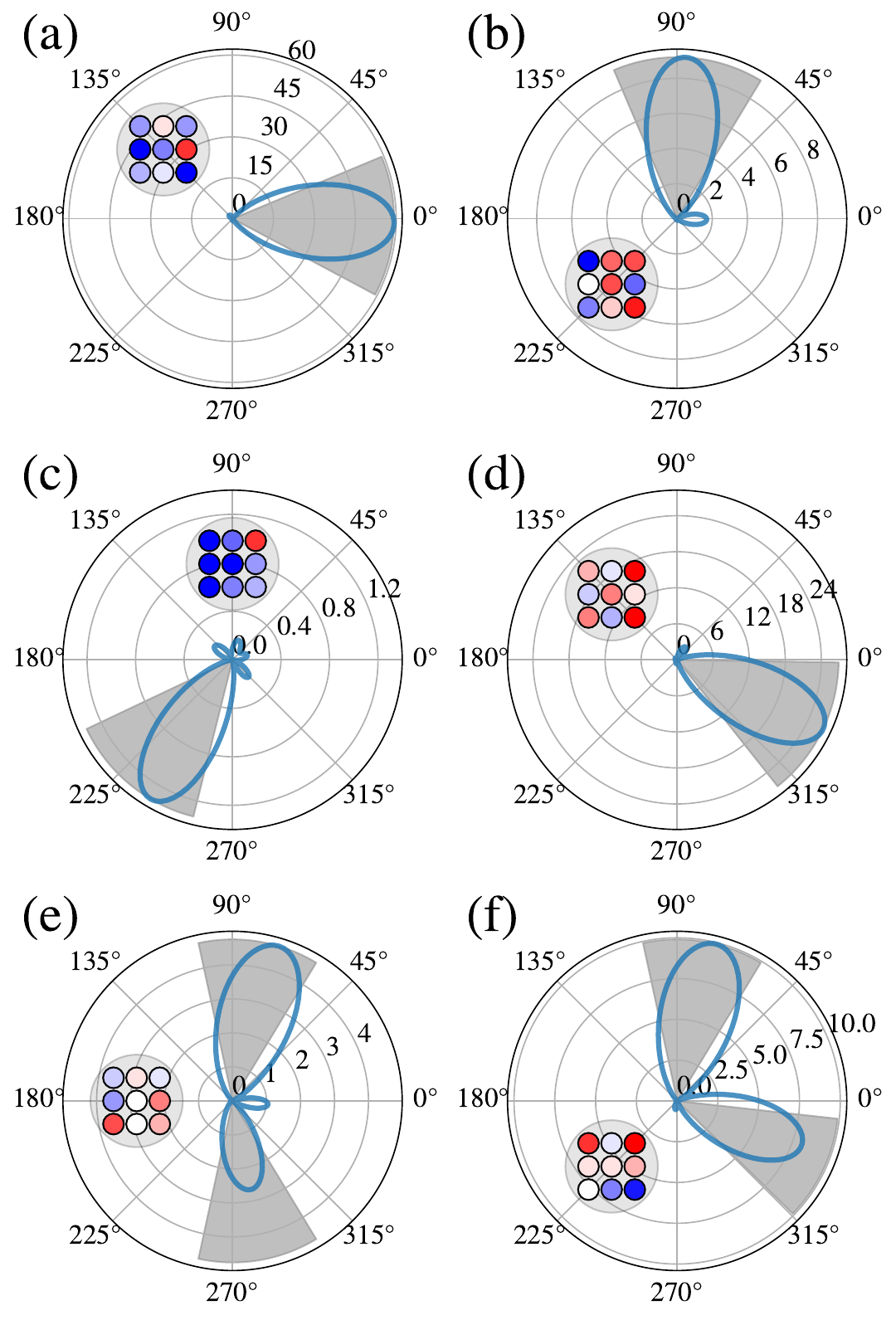}
	\caption{Far-field scattering current ($rj^{\text{sc},r}/R$) shown in polar coordinates and corresponding TDP configurations obtained by DE for different target angle ranges, shown with shaded areas to be maximized, over 200 generations.
(a)-(d) show single target angle ranges, while (e)-(f) display two angle ranges as simultaneous targets. The color of the dots in the TDP indicates their potential values, with the matching color map shown in Fig.~\ref{fig1}(a). }
    \label{fig3}
\end{figure}

To assess whether extended optimization yields improved performance or reveals alternative high-fitness configurations, the DE algorithm was continued for a total of 500 generations.
This extended run is shown by the dashed curves in Fig.~\ref{fig2}(a), with the higher final fitness after 500 generations achieved by a more pronounced peak within the target range.
The right insets of Fig.~\ref{fig2}(a) compare the optimal configurations after 200 and 500 generations (bottom and top, respectively), which differ significantly in their composition. 
In both cases, the target is achieved with a complex arrangement of dot potentials that would be difficult to determine without the optimization process. 
The richness of the solution space is demonstrated by the diverse independent solutions discovered for the same target. 
The electron density and total current flow in the near-field are mapped in Fig.~\ref{fig2}(b), and reveal a very strong current beam in the desired direction. 
This strong correspondence between near- and far-field scattering is not universal, and can be masked by interference with additional notes provided in the Supporting Information.

Figs. \ref{fig3}(a)-(d) illustrate that this approach also successfully finds suitable configurations for other target angles, demonstrating that the 9-dot TDP can controllably vary the deflection of an electron beam. 
The inset in each case shows the corresponding color-coded TDP optimal configuration, and the corresponding near-field maps are given in the Supporting Information.
In each case, a significant peak in the scattered current is achieved in the target range, where we note that the overall magnitude is greater for forward directions due to the general suppression of backscattering in graphene~\cite{katsnelson2006chiral,heinisch2013mie}.
In Figs.~\ref{fig3}(e),(f), we extend the approach to multiple angular ranges, allowing us to create designer beam splitters. 
This is achieved by modifying the first term in the fitness function in Eq.~\eqref{eq.optim} to account for two target angular ranges with details provided in the Supporting Information.

\paragraph{Discussion.}
The results so far demonstrate the capability of the TDP to adjust the direction and intensity of a scattered electron beam in the far-field.
This enables selective beam splitting and redirection in arbitrary directions, and has immediate applications in electron optics.
However, both the system geometry and the optimization procedure can be generalized to other physical quantities or target behaviors.  

\begin{figure}	\centering
	\includegraphics[scale=0.55]{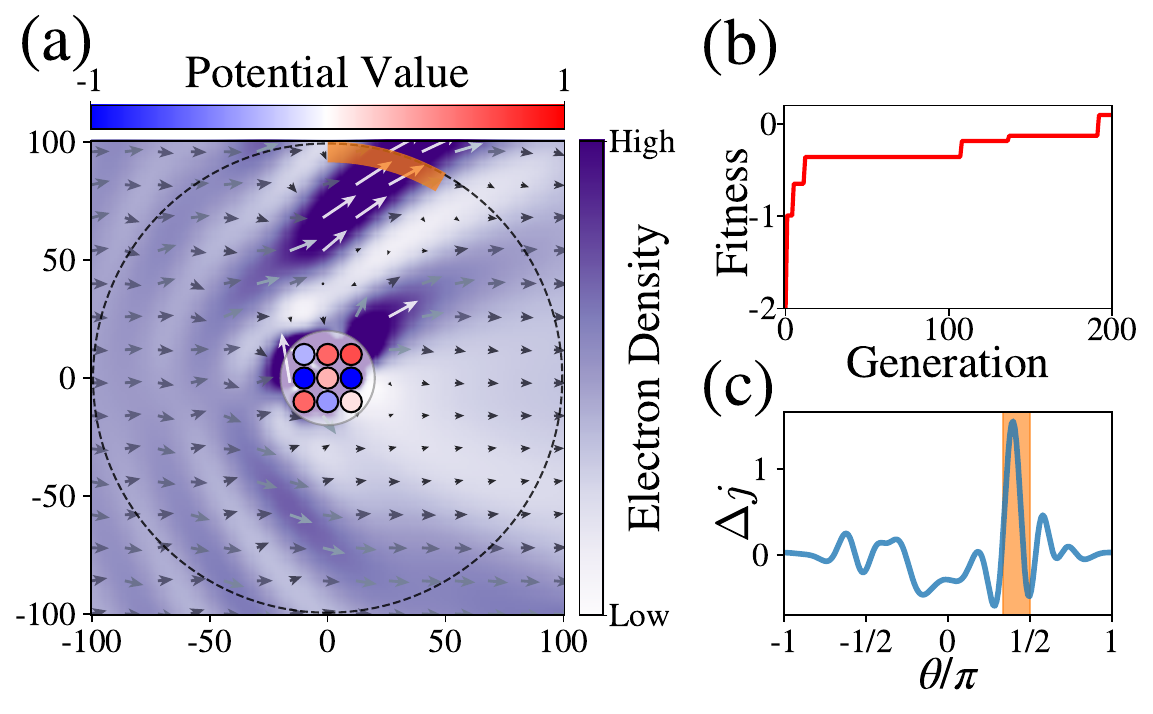}
	\caption{Optimization of near-field current using DE.
(a) Electron density (background color) and current flow (arrows) for indicated optimized TDP, where the color of each dot represents its potential value.
(b) Fitness improvement over 200 generations.
(c) Optimized $\Delta \bm{j}$ versus angle along the dotted black circle in (a), with the shaded region indicating the target angles. }
    \label{fig4}
\end{figure}

In small devices, or where it is important to locally regulate current flow in a specific region, an optimization of the near-field behavior may be more relevant.
This is the case, for example, if the deflected current is to be detected by a probe positioned near the scattering region.
To perform near-field optimization, we replace the far-field current in the fitness function with the position-dependent difference between the total and incident currents, $\Delta \bm{j(r,\theta)}=\bm{j(r,\theta)}-\bm{j}_\text{inc}\bm{(r,\theta)}$, where the calculation of the total current $\bm{j(r, \theta)}$ uses both incident and scattered wavefunctions.
This accounts for both the scattering and the interference between the scattered and incident waves.

Fig.~\ref{fig4}(a) shows a TDP with local current optimized for both a target angular range $[\pi/3, \pi/2]$ (orange shading) and a fixed distance $r=100$ (dotted circle) from the center of the TDP. 
After 200 generations of optimization (Fig.~\ref{fig4}(b)), $\Delta \bm{j}$ has a pronounced peak within the target range in Fig.~\ref{fig4}(c), corresponding to the significant local current and electron density at the desired location in Fig.~\ref{fig4}(a). 
Again we note that the sharply directed, asymmetric current achieved here cannot be achieved with a single dot, and relies on the interplay of the independently tuned dots in the TDP which enable constructive interference and rich, angle-selective patterns.

Our approach can also be generalized to different dot geometries or incident wave profiles. 
We have focused on a regime where the dots, TDP and Fermi wavelength are of the same order of magnitude and the TDP acts as a single, tunable effective scatterer.
By increasing the radius, separation and/or carrier density, we can operate instead in the optical regime where electrons scatter consecutively from different dots and where regular lattices have been shown to display robust experimental signatures~\cite{caridad2016electrical}. 
To maximize the proportion of the electron beam that is scattered and minimize interference effects, realistic electron optic devices are also more likely to employ focused beams instead of the simple incident plane wave considered here~\cite{boggild2017two, zhao2023electron, wan2021dirac}.

From a realization perspective, using individually controllable gates to define a TDP
enables each dot to be independently tuned \emph{in situ}, similar to the experimental setups already widely employed to investigate confined states and qubit behaviour in QDs~\cite{banszerus2018gate,freitag2016electrostatically,recher2010quantum,ge2020visualization}.  
This allows a single TDP to be a multi-purpose electronic component which can be fully reconfigured for different functionalities as required.
Alternatively, dielectric patterning~\cite{forsythe2018band, barcons2022engineering} can be used to vary the potential profile using only a single gate. 
In either case, extending the approach to multi-layer systems and dual-gated architectures gives access to additional control and functionalities.
For example, biased regions in bilayer graphene introduce valley-dependent current flows~\cite{solomon2021valley}, which could be harnessed for tunable valley-dependent beam splitters and filters in optimized TDP devices.

To conclude, the tunable dot platform enables precise control of electron flow in graphene.
Independently tuned dot potentials offer an exceptional flexibility to tailor the scattering of an incident electron wave.
We demonstrated that differential evolution is an effective way to determine the dot configuration required for specific target functionalities, such as beam deflection and splitting, in both the near- and far-field regimes. 
The programmability and reconfigurable nature of the TDP position it as a versatile and powerful platform for future graphene-based electronic and electron optic devices.


\begin{acknowledgement}
The authors acknowledge funding from the Irish Research Council under the Laureate awards and the Government of Ireland Postdoctoral Fellowship Program.  
\end{acknowledgement}


\providecommand{\latin}[1]{#1}
\makeatletter
\providecommand{\doi}
  {\begingroup\let\do\@makeother\dospecials
  \catcode`\{=1 \catcode`\}=2 \doi@aux}
\providecommand{\doi@aux}[1]{\endgroup\texttt{#1}}
\makeatother
\providecommand*\mcitethebibliography{\thebibliography}
\csname @ifundefined\endcsname{endmcitethebibliography}  {\let\endmcitethebibliography\endthebibliography}{}


\newpage
\appendix



\section*{\centering{Supporting Information}}



\maketitle

\setcounter{equation}{0}
\setcounter{figure}{0}
\setcounter{table}{0}
\setcounter{page}{1}
\makeatletter
\renewcommand{\theequation}{S\arabic{equation}}
\renewcommand{\thefigure}{S\arabic{figure}}
\renewcommand{\bibnumfmt}[1]{[S#1]}
\renewcommand{\citenumfont}[1]{S#1}

This supplemental material contains further detail on the scattering and optimization methodologies used in this work, as well as additional and complementary results to those presented in the main paper.

\section{Details of the scattering theory}\label{appendix1}
Following the approach in Ref. ~\cite{heinisch2013mie}, the incident Dirac plane wave with energy $E = \eta_0 k$ reads:
\begin{equation}
\psi_\text{inc} = \frac{e^{i\bm{k} \cdot \bm{r}} }{\sqrt{2}} 
\begin{pmatrix} 
1 \\ 
\eta_0  
\end{pmatrix}, 
\end{equation}
where $\eta_0 = \pm 1$ is the band index for conduction or valence bands. 
According to Eq. (1) in the main text, the incident wave is expressed in terms of the Bessel functions $J_m$ \cite{heinisch2013mie,heinisch2015electron}:

\begin{equation}
\psi_\text{inc}(kr, \theta) = \frac{1}{\sqrt{2}} \sum_{m=-\infty}^{\infty} i^m e^{im\theta} 
\begin{pmatrix} 
J_m(kr) \\ 
i\eta_0 e^{i\theta} J_{m+1}(kr) 
\end{pmatrix}.
\end{equation}
The scattered and transmitted wavefunctions are also expressed in terms of the Hankel function of the first kind, $H_m^{(1)}$, and the Bessel function $J_m$, respectively as:
\begin{equation}
\psi_{\text{sc}}(kr, \theta) = \frac{1}{\sqrt{2}} \sum_{m=-\infty}^{\infty} C^{\text{sc}}_m i^m e^{im\theta} 
\begin{pmatrix} 
H_m^{(1)}(kr) \\ 
i\eta_0 e^{i\theta} H_{m+1}^{(1)}(kr) 
\end{pmatrix}, 
\end{equation}

\begin{equation}
\psi_{\text{tr}}(kr, \theta) = \frac{1}{\sqrt{2}} \sum_{m=-\infty}^{\infty}  C^{\text{tr}}_m i^m e^{im\theta} 
\begin{pmatrix} 
J_m(qr) \\ 
i\eta e^{i\theta} J_{m+1}(qr) 
\end{pmatrix}. 
\end{equation}
\begin{figure}
	\centering
	\includegraphics[scale=0.45]{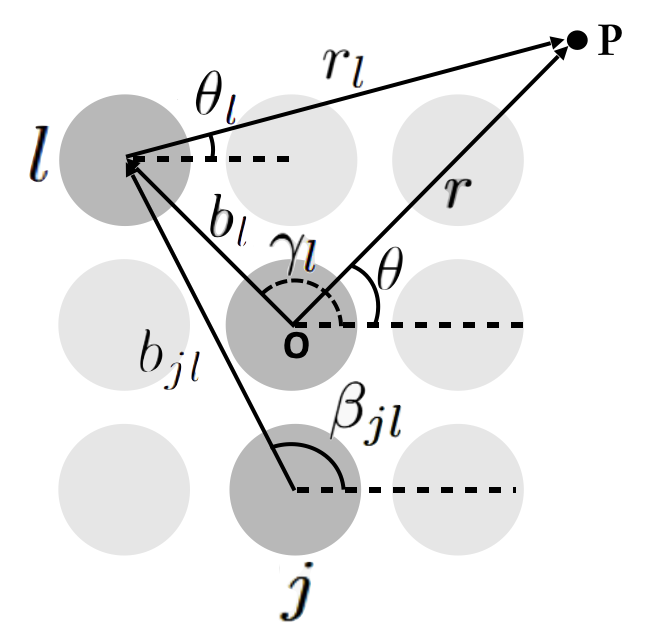}
	\caption{The framework for studying the TDP, illustrating the relevant vectors and angles. }
    \label{framework}
\end{figure}
The continuity of the wavefunction at the dot edges $r = R$ gives single dot coefficients as:
\begin{equation}
 C^{\text{sc}}_m = \frac{\eta_0  J_m(qR)J_{m+1 }(kR)-\eta  J_m(kR)J_{m+1 }(qR)}{\eta  H_m^{(1)}(kR)J_{m+1 }(qR)-\eta_0  H_{m+1 }^{(1)}(kR)J_m(qR)}, \nonumber
\end{equation}
\begin{equation}
 C^{\text{tr}}_m  = \frac{\eta_0J_m(kR)H_{m}^{(1)}(kR)-\eta_{0}J_m(kR)H_{m+1 }^{(1)}(kR)}{\eta  H_m^{(1)}(kR)J_{m+1 }(qR)-\eta_0  H_{m+1 }^{(1)}(kR)J_m(qR)}. 
\end{equation}

To extend to multiple dots, following Ref.~\cite{sadrara2019dirac},  the position of the $l$th dot is indicated by vector $\bm{b}_l$ (with corresponding angle $\gamma_l$) from the origin $O$, and a vector connecting the $j$th dot to the $l$th dot is represented by $\bm{b}_{jl}$ (with corresponding angle $\beta_{jl}$), as shown in Fig.~\ref{framework}. 
For simplicity and to maintain symmetry around $\theta = 0$, we consider the origin to be at the center of the TDP. 
Since the designed TDP consists of 9 dots, the origin is  positioned at the center of the central dot.
A given point $P$ in position $\bm{r}$ (with corresponding angle $\theta$) in the global polar coordinates has position $\bm{r}_l$ (with corresponding angle $\theta_l$) in the reference frame of $l$th dot, so that $\bm{r} = \bm{r}_l + \bm{b}_l$. 
Consequently, the incident wavefunction is written using
\begin{equation}
e^{i\bm{k} \cdot \bm{r}} = e^{i\bm{k} \cdot (\bm{b}_l + \bm{r}_l)} = e^{ikb_l \cos\gamma_l + ikr_l \cos\theta_l},\nonumber
\end{equation}
as
\begin{equation}
\psi_\text{inc}(kr_l, \theta_l) = \!\!\!\!\sum_{m=-\infty}^{\infty} \!\!\!\frac{i^m e^{ikb_l \cos\gamma_l} e^{im\theta_l}}{\sqrt{2}} \!
\begin{pmatrix} 
J_m(kr_l) \\ 
i\eta_0 e^{i\theta_l} J_{m+1}(kr_l) 
\end{pmatrix}.
\end{equation}

The scattered wavefunction at point $P$ originating from all individual dots is given by:  
\begin{equation}
\psi_{\text{sc}}(kr, \theta) = \psi_{\text{sc},l}(kr_l, \theta_l) + \sum_{j \neq l}^{N} \psi_{\text{sc},j}(kr_j, \theta_j).
\end{equation}
Here, the first term represents the contribution of the $l$th dot, while the second term accounts for the contributions from all other $j$th dots, all written in their own reference frame. Substituting the corresponding scattered wavefunctions yields
\begin{equation}
\begin{split}
\psi_{\text{sc}}(kr, \theta) &= \sum_{m=-\infty}^{\infty} \frac{C^{\text{sc}}_{ml} i^m e^{im\theta_l}}{\sqrt{2}} 
\begin{pmatrix} 
H_m^{(1)}(kr_l) \\ 
i\eta_0 e^{i\theta_l} H_{m+1}^{(1)}(kr_l) 
\end{pmatrix} \\
&\quad + \!\!\sum_{j \neq l}^{N} \sum_{m=-\infty}^{\infty} \!\!\!\!\frac{C^{\text{sc}}_{mj} i^m e^{im\theta_j}}{\sqrt{2}} \!\!
\begin{pmatrix} 
H_m^{(1)}(kr_j) \\ 
i\eta_0 e^{i\theta_j} H_{m+1}^{(1)}(kr_j) 
\end{pmatrix},
\end{split}
\label{eq:sclj}
\end{equation}
where $C^{\text{sc}}_{ml}$ and $C^{\text{sc}}_{mj}$ indicate the $m$-dependent scattering coefficients for $l$th and $j$th dots, respectively. To proceed, a consistent reference frame is required. Therefore, the second term of the scattered wave in Eq.~\eqref{eq:sclj} is rewritten in the reference frame of the $l$th dot using Graf's addition theorem \cite{meng2016bounds}. For $r_l < b_{jl}$, the theorem states
\begin{equation}
H_m^{(1)}(kr_j) e^{im\theta_j} =\!\!\!\! \sum_{n=-\infty}^{\infty} H_{m-n}^{(1)}(kb_{jl}) e^{i(m-n)\beta_{jl} + in\theta_l} J_n(kr_l).
\end{equation}
Thus, the scattered wavefunction can be expressed as:
\begin{equation}
\psi_{\text{sc}}(kr_l, \theta_l) = \psi_{\text{sc},l}(kr_l, \theta_l) + \sum_{j \neq l}^{N} \psi_{\text{sc},j}(kr_l, \theta_l).
\end{equation}
Finally, with $C^{\text{tr}}_{ml}$ representing the $m$-dependent transmission coefficient for $l$th dot, the transmitted wave within this dot is given by 
\begin{equation}
\psi_{\text{tr},l}(qr_l, \theta_l) = \sum_{m=-\infty}^{\infty} \frac{C^{\text{tr}}_{ml} i^m e^{im\theta_l}}{\sqrt{2}} 
\begin{pmatrix} 
J_m(q_l r_l) \\ 
i\eta_l e^{i\theta_l} J_{m+1}(q_l r_l) 
\end{pmatrix}.
\end{equation}
Substituting the wavefunction expansions above into the continuity equation for $l$th dot 
\begin{equation}
\psi_\text{inc}(kR_l, \theta_l) + \psi_{\text{sc}}(kR_l, \theta_l) 
= \psi_{\text{tr},l}(qR_l, \theta_l), 
\end{equation}
gives a system of equations with a coefficient matrix of size $MN \times MN$. This is solved numerically to give the scattering and transmission coefficients for each dot and mode. 

To obtain the radial component of the far-field scattered current we express the radial Pauli matrix as $\sigma_r = \sigma_x \cos \theta + \sigma_y \sin \theta$. The scattered current in the far-field limit ($r \rightarrow \infty$) is then given by:
\begin{equation}
j^{\text{sc},r}(r, \theta) = \psi_{\text{sc}}^{\dagger} \sigma_r \psi_{\text{sc}}.
\nonumber
\end{equation}
To evaluate this expression, the total scattered wavefunction in Eq.~\eqref{eq:sclj} must first be rewritten in the reference frame of the $l$th dot, once again, but for $r_l > b_{jl}$, using Graf's addition theorem, leveraging the asymptotic behavior of the Hankel functions at $r \rightarrow \infty$
\begin{equation}
H_m^{(1)}(kr_j) e^{im\theta_j} =\!\!\!\! \sum_{n=-\infty}^{\infty} \!\! J_{m-n}(kb_{jl}) e^{i(m-n)\beta_{jl} + in\theta_l} H_n^{(1)}(kr_l).
\label{eq:16}
\end{equation}
The far-field scattering current is then obtained as
\begin{equation}
j^{\text{sc},r}(r \rightarrow \infty, \theta) = \frac{2\eta_0}{\pi kr} \sum_{m,n=-\infty}^{\infty} \mathscr{S}_m \mathscr{S}_n^* e^{i(m-n)\theta},
\label{eq:17}
\end{equation}
where $\mathscr{S}_m$ is the effective scattering coefficient of the entire TDP, which is defined as
\begin{equation}
\mathscr{S}_m = \sum_{j=1}^{N} \sum_{n=-\infty}^{\infty} J_{n-m}(kb_j) e^{i(n-m)(\gamma_j + \pi)} i^{n-m} C^{\text{sc}}_{nj}.
\label{eq:18}
\end{equation}

It is important to note that we typically work in the low-energy regime, where lower-order modes dominate the scattering process. 
A few low-order modes are sufficient to calculate the scattering properties. 
At higher energies, the contributions of higher-order modes become more significant, resulting in complex interference patterns. Similarly, real space mapping of the currents and densities over larger regions can require higher-order modes to sufficiently represent the incident plane wave.

\section{Comparison of single dot and TDP scattering}\label{appendix2}
\begin{figure*}
	\centering
	\includegraphics[width = 1\linewidth]{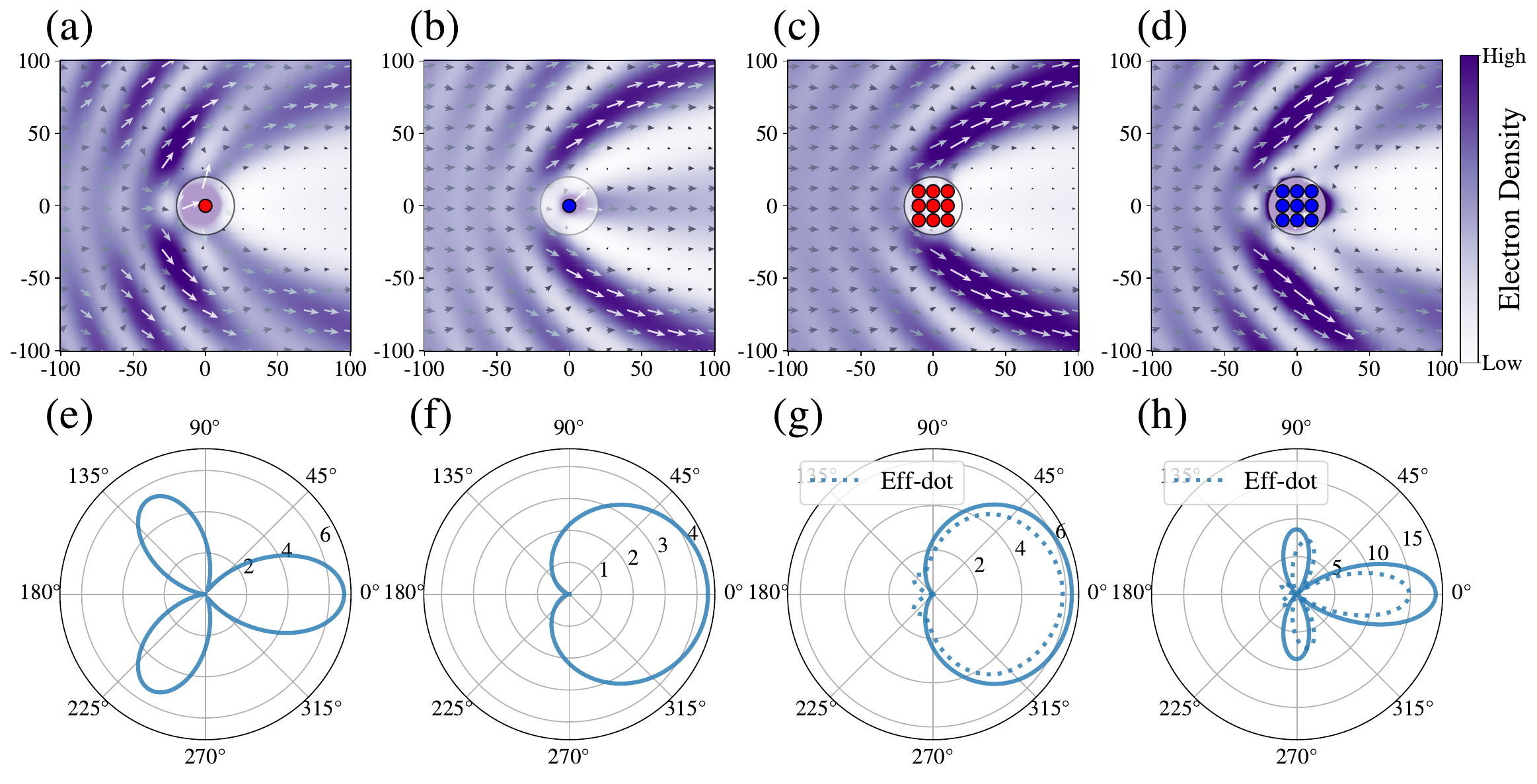}
	\caption{ Top panels: Electron density (background color) and probability current (arrows) in the near-field region for (a) a single QD with $V = 1$ (red dot), (b) a single QD with $V = -1$ (blue dot), (c) a TDP composed of identical QDs with $V = 1$, and (d) a TDP with $V = -1$ where $R = 4.5$ and $E = 0.1$. Bottom panels: Far-field scattering current corresponding to the  configurations in the top panels. The dotted curves in (g) and (h) represent effective QDs with $R_{eff} = 19.2$, $V = 1$ and $R_{eff} = 18.7$, $V = -1$, respectively.  }
    \label{fig2s}
\end{figure*}
Fig.~\ref{fig2s} presents additional near- and far-field results to complement the single dot and simple uniform TDP results from Fig. 1(b) of the main paper. 
Panels (a)–(d) show the electron density (background color) and near-field current flow (arrows) for QDs and TDPs, with the relevant dot system shown in each case.
Interference between the incident and scattered waves creates distinct peaks and dips in the electron density near the QDs. Regions of higher electron density correspond to stronger current flow. 
Due to the inherent symmetry of the system, the scattering patterns are symmetric with respect to the $x$-axis in all cases. 
For TDPs with equal potential values $V = 1$ or $V = -1$ on all dots, the scattering patterns differ significantly from those of single QDs with the same potential.

Figs.~\ref{fig2s}(e)–(h) present the angular dependence of the far-field scattered current corresponding to the configurations in (a)–(d). 
While a single QD with $V = 1$ exhibits three preferred scattering directions, the same structure with $V = -1$ results in a single, broad forward scattering peak around $\theta = 0$.
The far-field scattering currents of the TDPs in Figs.~\ref{fig2s}(g) and (h) show behaviors markedly different from that of single QDs with the same potentials in Figs.~\ref{fig2s}(e) and (f).  
As discussed in the main paper, a uniform TDP acts like one effective large dot, as indicated by the far-field behavior in Figs.~\ref{fig2s}(g) and (h), where the dotted curve represents the result for a larger single dot with the same potential. 
The similar behavior of single large dots and TDPs, and the markedly different behavior of single small dot, highlights the effect of the overall size of the scattering region is more significant than the structure of the TDP.

The near- and far-field results do not always have an immediate and intuitive correspondence with each other.
For example, the enhanced forward far-field scattering in Fig.~\ref{fig2s}(e) is not evident in the vicinity of the dot in Fig.~\ref{fig2s}(a), due to a shadow effect immediately behind the dot. 
However, strong right-going current is observed towards the top and bottom of the mapped region 
Similarly, interference with the incident wave can disguise strong scattering, particularly in regions in front of the dot. 

We also note that in the main paper, the TDP parameters are presented in a scaled form.
As the energy and length scales enter together in terms such as $kR$, there are many possible physical realizations.
For example, with $E \approx 14.6  \text{meV}$ and $V \approx 146  \text{meV}$, dots of radius $R = 4.5 \text{nm}$ would be required.
These values fall within the tolerance range of current experimental fabrication techniques \cite{bai2018generating,gutierrez2016klein, zhao2023electron, wan2021dirac}, demonstrating the feasibility of realizing this platform as a practical component.
However the exact behavior can be reproduced, for example, by scaling the dot radius up and the energy and potential values down by the same factor.

\section{Optimization using Differential Evolution}\label{appendix3}

Optimization begins by generating $n_p$ random candidate solutions from a uniform distribution, then mapping them to the discrete potential levels allowed, to form the initial population.
Each individual is represented by a vector $\mathbf{x}_i$ for $i = 1, 2, \ldots, n_p$. 
The vector $\mathbf{x}_i = [x_{i1}, \ldots, x_{iN}]$ corresponds to a specific configuration of the TDP, with $x_{ij}$ representing the potential of the $j$th dot.
The fitness of each individual is computed using a defined fitness function such as Eq. (5) in the main text that evaluates how well a given configuration meets the desired criteria. 

Then, new candidate solutions are generated through mutation and recombination as follows \cite{price2006differential}.  
Corresponding to individual $\mathbf{x}_i$ in the population, a mutant vector $\mathbf{v}_i$ is built using  
\begin{equation}
\mathbf{v}_i = \mathbf{x}' _{r} +F  \left(\mathbf{x}^{''}_{r} - \mathbf{x}^{'''}_{r}\right),
\end{equation}
where $\mathbf{x}^{'}_{r},\mathbf{x}^{''}_{r}, \mathbf{x}^{'''}_{r}$ are distinct vectors randomly selected from the population, and $F$ is the mutation factor that controls the amplification of the differential variation. In our implementation, we fix $F = 0.3$.

Next, the trial vector $\mathbf{u}_i$  with elements $u_{ij}$ is created using a binomial crossover, combining elements  $v_{ij}$ from $\mathbf{v}_i$ and $x_{ij}$ from $\mathbf{x}_i$ as:
\begin{equation}
u_{ij} =
\begin{cases} 
v_{ij}, & \text{if } \textrm{rand}_j \leq c_r \\ 
x_{ij}, & \text{otherwise}, 
\end{cases}
\end{equation}
where $c_r$ is the crossover probability and $\textrm{rand}_j$ is a random number uniformly drawn from the interval $[0, 1]$.
In the event that no crossover happens, to ensure genetic diversity, we constrain at least one component to come from the mutant vector during crossover. While mutation encourages exploration of new possibilities, crossover emphasizes refining the current solutions based on existing good solutions. To balance the search around well-established solutions, we set the crossover probability to decay over generations as $c_r(g)=c_r(0)\times(1-g/N_g)$, where $g$ stands for generation and $N_g$ is the total number of generations. A higher crossover rate in early generations promotes diversity by allowing more mutant components to enter the trial vector. As the population converges, a lower crossover rate reduces disruptive changes, preserving favorable structures while still allowing limited mixing. This strategy aims to support broad search in early stages and gradual refinement in later ones.

In the next step, selection as a deterministic rule is performed, where the fitness of the trial vector $\mathbf{u}_i$ is compared to that of the parent vector  $\mathbf{x}_i$. 
If the trial vector has a higher fitness score (specified as a better solution), it replaces the parent in the next generation; otherwise, the parent is retained. 
This ensures that each generation maintains or improves previous solutions, while preventing the loss of high-quality configurations. 
This process iteratively continues over multiple generations, progressively guiding the population toward optimal TDP configuration, until a defined criterion is met \cite{price2006differential,storn1997differential}. 
In our case, we set a fixed maximum number of generations as the stopping condition.

\section{Fitness functions}

The fitness function used to optimize the far-field current within a single target angular range is defined in Eq. (5) of the main text. 
It consists of two competing terms: a positive term that promotes the enhancement of current with the weighting factor $W=0.4$ within the desired angular range, and a negative term that penalizes current with the weighting factor $(1-W)$ outside this range. The weighting factors that are used to control the relative importance of enhancing the current within the target angular regions and suppressing it outside can be tuned based on the width of the corresponding angular intervals to ensure balanced optimization performance.

As shown in the inset of Fig. 2(a), negative fitness values in early generations indicate that large currents outside the target region still dominate, but as the optimization progresses toward the desired current distribution, the fitness becomes positive as its magnitude is now determined by the strength of scattering in the target angular range. 

When the optimization aims to maximize current within two distinct angular ranges $\theta_{\text{in},1}$ and $\theta_{\text{in},2}$, as illustrated for the beam splitting examples in Fig. 3(e),(f) of the main text, the fitness function is extended as follows:
\begin{equation}
\text{Fitness} = \sqrt{J_1J_2}- W_\text{out} \int_{\theta_{\text{out}}} j^{\text{sc},r}(\theta) d\theta,
\label{eq:fitnesssplitter}
\end{equation}
where, 
\begin{equation}
J_1=W_\text{in,1}\int_{\theta_{\text{in},1}} j^{\text{sc},r}(\theta)  d\theta ,~~~ J_2=W_\text{in,2}\int_{\theta_{\text{in},2}} j^{\text{sc},r}(\theta)  d\theta. \nonumber
\end{equation}
Here, we set the weighting factors as $W_{\text{in},1}=W_{\text{in},2}=1.2$, and $W_{\text{out}}=W_{\text{in},1}/2$ to place greater emphasis on the maximization term relative to the minimization term.
The product in Eq.~\eqref{eq:fitnesssplitter} rewards solutions that produce scattered currents in both desired directions over those with a large scattered current in only one of the target ranges.

Analogously to the far-field case, a corresponding fitness function can be formulated for near-field current control as:
\begin{equation}
\text{Fitness} = W \int_{\theta_{\text{in}}} \Delta j (r,\theta) \, d\theta - (1-W) \int_{\theta_{\text{out}}} \Delta j^+ (r,\theta) \, d\theta,
\end{equation}
where $\Delta \bm{j}(r, \theta) = \bm{j}(r, \theta) - \bm{j}_{\text{inc}}(r, \theta)$ denotes the difference between the total and incident currents, as defined in the main text. 
The integration is performed along a fixed radial distance from the scatterer, as shown by a dotted circle in Fig. 4(a)) of the main text. 
With the weighting factor $W = 0.4$, the first term involving $\theta_{\text{in}}$ refers to the angular region where the current is to be maximized -- corresponding to the highlighted segment in Fig. 4(a),(c). 
The second term, as before, aims to suppress unwanted current outside the target region, however for the near-field case this term only considers the positive (outward) component of the current, denoted by $\Delta j^+$.
This ignores any inward (negative) contributions, which do not arise in the far-field case.

\section{Near-field currents for different optimization targets}\label{appendix3}

\begin{figure}	\centering
	\includegraphics[width = 0.6\linewidth]{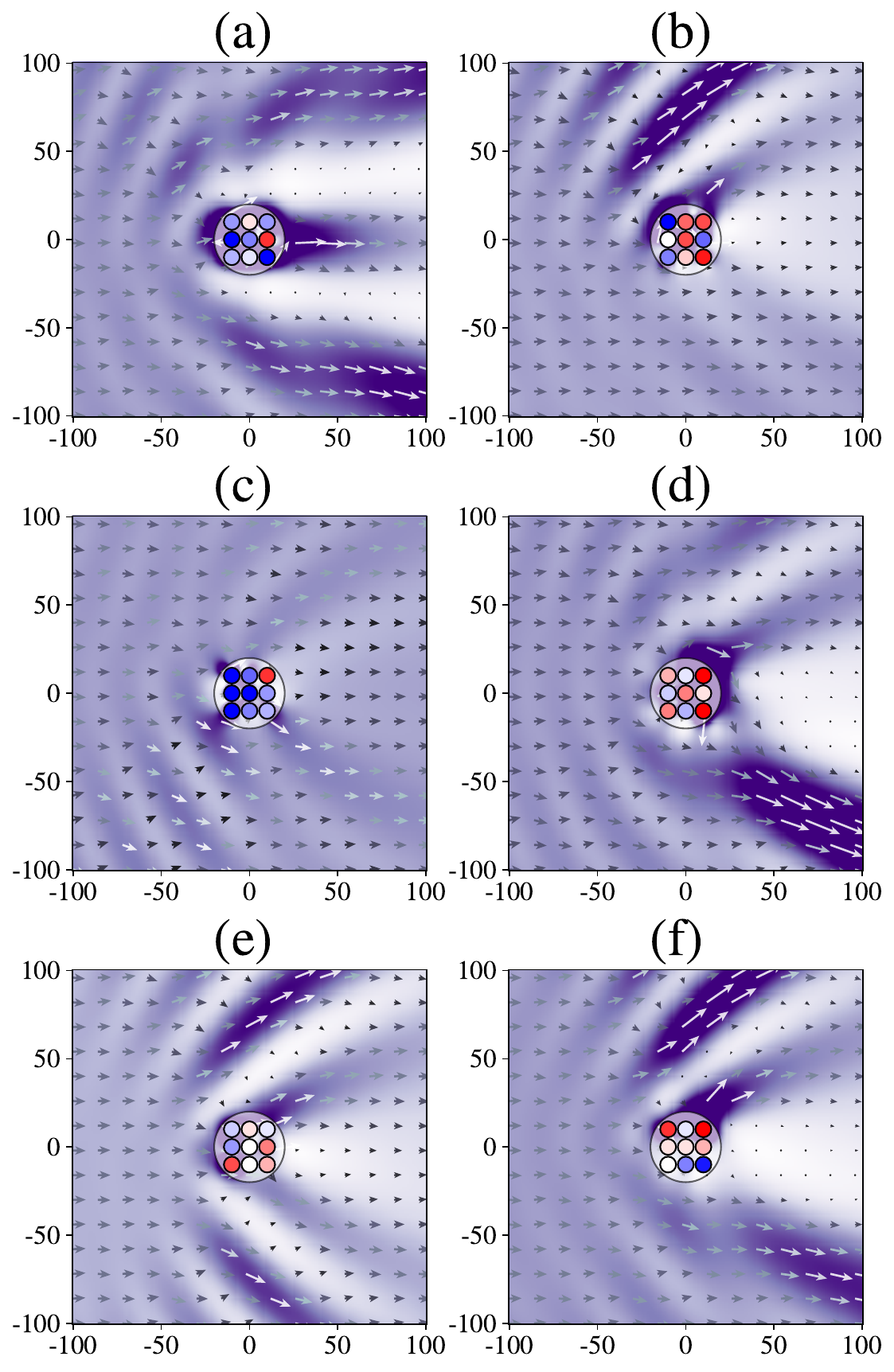}
	\caption{ Electron density (background color, with color map as in previous figures) and near-field current (arrows) corresponding to TDP configurations optimized by DE for the different target angle ranges shown in Fig. 3 of the main text, shown in the same order. Dot colors indicate their potential values, following the color map in Fig. 1(a) in the main text. } 
    \label{fig3s}
\end{figure}

 Fig.~\ref{fig3s} shows the electron density and the near-field current corresponding to the optimal TDP configurations obtained for the different targets considered in Fig. 3 of the main text, with the panels presented in the same order.
 As seen here and also previously in Fig.~\ref{fig2s}, there is not always a direct or intuitive correspondence between near-field and far-field behaviors. 
 For instance, panel (a) in both Fig. 3 of the main paper and Fig.~\ref{fig3s} here again illustrate this disconnect. 
 Therefore, optimizing the far-field scattered current may not be suitable in scenarios where the deflected current is intended to be detected by a probe positioned close to the scattering region, and instead the near-field optimization discussed in the main text may be more appropriate.

\newpage

\end{document}